\documentstyle[aps,prb,twocolumn]{revtex}
\begin{document}
\title{Spin-wave excitation spectra and spectral weights \\
in square lattice antiferromagnets}
\author{Rajiv R. P. Singh}
\address{Department of Physics, University of California, Davis,
California 95616}
\author{Martin P. Gelfand}
\address{Department of Physics, Colorado State University, Fort
Collins, Colorado 80523}
\date{\today}
\maketitle

\begin{abstract}
Using a recently developed method for calculating series expansions
of the excitation spectra of quantum lattice models, we
obtain the spin-wave spectra for square lattice, $S=1/2$
Heisenberg-Ising antiferromagnets.
The calculated spin-wave spectrum for
the Heisenberg model is close to but noticeably different from
a uniformly renormalized classical (large-$S$) spectrum with the
renormalization for the spin-wave velocity of approximately $1.18$.
The relative weights of
the single-magnon and multi-magnon contributions to neutron scattering
spectra are obtained for wavevectors throughout the Brillouin zone.
\end{abstract}
\pacs{PACS:75.10.Jm,75.30.Ds,74.25.Ha}

\narrowtext

Thermodynamic properties and excitation spectra of two-dimensional
quantum antiferromagnets have attracted much attention, especially
because of their potential relevance to high temperature
superconductivity in the cuprate perovskites.\cite{CHN} Methods
based on high-order series expansions about the Ising model
have proven to be very successful in
accurate calculations of the thermodynamic properties of
two-dimensional antiferromagnets.\cite{square,WOH}
One major limitation of these methods
has been their inability to deal directly with dynamical
properties or excitation spectra. These quantities have so far been
studied, within the series approach, via frequency moments\cite{SFLS}
and single mode approximations.\cite{SMA}  The reliability of using
only a few moments to reproduce spectral lineshapes has come into
question, for example, in the case of the two-magnon Raman
spectra.\cite{canali}
Currently efforts are underway to use a large number of frequency
moments to obtain the dynamical properties by numerical
analytical continuation.\cite{sokol}

Recently one of us\cite{gelfand} has shown that a
series expansion method can be used to directly calculate
excited state properties of quantum many-body systems.
Here, we apply this method to calculate the energies and
spectral weights of the elementary excitations
in square lattice, $S=1/2$ Heisenberg-Ising magnets
by expansion around the Ising limit to order $(J_\perp/J_z)^{10}$.

Let us briefly summarize the results.
The spin-wave spectrum has an energy-gap for $J_\perp<J_z$,
which vanishes
as the Heisenberg limit ($J_\perp=J_z$) is reached.
Using series extrapolation methods to estimate the
dispersion in the Heisenberg limit,
we find a spin-wave velocity
which agrees with $1/S$ expansions and other previous estimates.
Along the line $q_x=q_y$, the dispersion
is nearly uniformly renormalized
with respect to its classical value.
Measurable deviations are found in other directions.
In particular, along the line $q_x+q_y=\pi$
the spin-wave energy is maximized at
$(\pi/2,\pi/2)$ and it exceeds its value at $(\pi,0)$ by about 7\%.

Another quantity that has been calculated by this
method is the single-magnon contribution to neutron scattering,
that is, the coefficient associated with the delta-function
in the dynamic structure factor.
When compared with the equal-time transverse correlation function,
it yields the relative
weight of multi-magnon and single-magnon contributions to the
neutron scattering spectra  at different wavevectors.
We find that over substantial parts of the Brillouin zone
approximately 20\% of the total spectral weight
is associated with the multi-magnon excitations even though
the relative weight of such excitations
vanishes near $(0,0)$ and $(\pi,\pi)$.

The Heisenberg-Ising Hamiltonian under consideration is defined by
\begin{equation}
{\cal H}=J_z\sum_{\langle i,j \rangle}
S_i^z S_j^z+\alpha(S_i^xS_j^x+S_i^yS_j^y),
\end{equation}
where the sum runs over nearest-neighbor pairs on a square
lattice for which the lattice constant is the unit length
($a\equiv 1$), and $\alpha=J_\perp/J_z$.
In the Ising limit, $\alpha=0$, there are two degenerate ground states
and the single-magnon excitations are single spin-flips with respect
to the N\'eel states. In this limit the excitations are purely local,
or, alternatively,
one could say that the magnon energies are degenerate over the entire
band. For $\alpha\ne 0$, the single-magnon states evolve into a set of
states with a nonzero dispersion.
The key to calculating the spin-wave dispersion
is to construct an effective Hamiltonian for the states
which are the natural, perturbatively constructed extensions
of the single spin-flip states at finite $\alpha$.
The effective Hamiltonian
is then readily diagonalized by Fourier transformation.

Because the spins on the two sublattices are oriented in
opposite directions, the single spin-flip states naturally
divide into two sets, those corresponding to $S^z=+1$ and those
with $S^z=-1$. The effective Hamiltonian, which conserves $S^z$,
thus connects only
the basis states with spin-flips on the same sublattice: this
ensures that the spin-wave spectrum is degenerate between
wavevectors $(q_x,q_y)$ and $(\pi-q_x,\pi-q_y)$. The full
effective Hamiltonian to order $\alpha^{10}$ is presented in
Table~\ref{spec_table}.

For $\alpha\ne 1$ there is a gap in the spectrum, the minimum
being at $(0,0)$ and $(\pi,\pi)$. The expansion
for the gap is
\begin{eqnarray}
2&-&(5/3)\alpha^2+0.31712963 \alpha^4-0.41923376 \alpha^6 \nonumber\\
&+&0.27099699 \alpha^8-0.38943351 \alpha^{10} +\cdots
\end{eqnarray}
which agrees completely with the ``mass gap''
calculated by Zheng et al. \cite{WOH}
General arguments tell us that the
gap must close in the Heisenberg limit, $\alpha=1$.
Moreover, we expect
that for small $q=|{\bf q}|$ the spectrum has the form
$\epsilon({\bf q})\approx[A(\alpha)+B(\alpha)q^2]^{1/2},$
where $A(\alpha)\to 0$ as $\alpha\to 1$.
The spin-wave velocity in the Heisenberg limit is given by
$B(1)^{1/2}$.
In order to calculate the spin-wave velocity,
we expand $\epsilon({\bf q})$ in powers of $q$,
$\epsilon (q)= C(\alpha) + D(\alpha)q^2 + \cdots,$
and identify $C=A^{1/2}$ and $D=B/2A^{1/2}$.
Thus the square of the
spin-wave velocity for the Heisenberg model is given by the
$\alpha\to 1$ limit of the series
\begin{eqnarray}
2C(\alpha)D(\alpha)=4&\alpha^2&-2.305555 \alpha^4+
           2.410512 \alpha^6 \nonumber\\
&-&3.064895 \alpha^8+4.100549 \alpha^{10}
+\cdots .
\end{eqnarray}
Since we expect only a weak, energy-like singularity in this series
at $\alpha=1$, we can sum it by Pad\'e approximants. The near-diagonal
approximants [2/2], [2/3] and [3/2] give estimates of
$2.779$, $2.785$ and $2.785$ respectively, which leads to values
for the spin-wave velocity $c/Ja$ of $1.667$, $1.669$ and $1.669$.
In the large-$S$ limit $c/Ja=\sqrt{2}$, so the quantum
renormalization of the spin-wave velocity is $Z_c \approx 1.18$.
This number is in excellent agreement with the high-order spin-wave
calculation \cite{Igarashi}
and previous indirect estimates \cite{rhos}
using the hydrodynamic relation $c^2=\rho_s/\chi_\perp$,
where $\rho_s$ is the spin-stiffness and $\chi_\perp$ the
uniform transverse susceptibility.

Away from the gapless points $(0,0)$ and $(\pi,\pi)$
the spin-wave spectrum for the Heisenberg
model can be estimated by direct Pad\'e approximants.
Along the line $q_x=q_y$,
the spectrum, within our numerical
uncertainties, is uniformly renormalized with respect to
the classical (large-$S$) spectrum. However, along the line
$q_x=0$ appreciable differences appear.
In particular, the excitation energy at $(\pi,0)$
appears to be a shallow local minimum
along this line, and
is lower than that at $(\pi/2,\pi/2)$ by about 7\%.
Note that at the classical and $1/S$ levels the spin-wave spectrum
is degenerate at $(\pi,0)$ and $(\pi/2,\pi/2)$.
A plot of the dispersion relation along selected directions
is shown in Fig.~\ref{disp_fig}.
A hint of the deviations from classical spin-wave theory which
we have found near $(\pi,0)$
can be seen in the results of the projector quantum Monte Carlo
calculations of Chen {\it et al.},\cite{ChenDingGoddard}
even though those authors emphasize the similarity
of the $S=1/2$ spin-wave spectrum and the uniformly renormalized
classical spectrum.
The first calculation of the spectrum by spin-wave expansion to
order $1/S^2$ by Igarashi and Watabe~\cite{IgarashiWatabe} also
suggested that $(\pi,0)$ would be a local minimum for the
spin-wave spectrum; however, that minimum is no longer present
in the more recent spin-wave calculations by Igarashi.\cite{Igarashi}

We now turn to the spin-wave spectral weights.
The magnetic neutron scattering cross section is proportional
to the dynamic structure factor, given by the expression
\begin{eqnarray}
\hat S({\bf q},\omega)=
\int dt e^{-i\omega t}
\sum_{\bf r}
e^{i{\bf q}\cdot{\bf r}}
\langle && S^x({\bf 0},0)S^x({\bf r},t) \nonumber \\
&&+S^y({\bf 0},0)S^y({\bf r},t)\rangle.
\end{eqnarray}
We consider the $T=0$ limit, where the angular brackets refer
to ground state expectation values.
In general, we expect $\hat S({\bf q},\omega)$ to consist of a sum
of two parts,
\begin{equation}
\hat S({\bf q}, \omega)= A({\bf q})\delta (\omega-\epsilon({\bf q}))
+ B({\bf q},\omega).
\end{equation}
Here $\epsilon ({\bf q})$ is the spin-wave dispersion and $A({\bf q})$
is the residue or spectral weight associated with the spin-waves.
$B({\bf q},\omega)$ is associated with multi-magnon excitations,
which are present because a single spin-flip in an antiferromagnet
cannot be exactly represented as a superposition of
single-magnon states.
Integrating $\hat S({\bf q}, \omega)$ over all frequencies yields
the equal-time correlation function
\begin{equation}
\hat S({\bf q})=\sum_{\bf r} e^{i{\bf q}\cdot {\bf r}}
\langle S^x({\bf 0},0) S^x({\bf r},0)+
        S^y({\bf 0},0) S^y({\bf r},0)\rangle.
\end{equation}
To determine the residue $A({\bf q})$, we need
to restrict the intermediate states that arise in the
calculation of the expectation values to be single magnon states,
which gives the expression
\begin{equation}
A({\bf q})=\sum_{\bf r} e^{i{\bf q}\cdot {\bf r}}
\langle S^x({\bf 0},0) {\cal P} S^x({\bf r},0)+
S^y({\bf 0},0) {\cal P} S^y({\bf r},0)\rangle,
\end{equation}
where ${\cal P}$ is the projection onto the manifold
of single-magnon states. In the cluster expansions, each
single-magnon state evolves with the coupling $\alpha$,
and is of the form
\begin{equation}
|\psi_i\rangle=|i\rangle+\sum_n C_{i,n}(\alpha)|n\rangle,
\end{equation}
where $|i\rangle$ is a single-magnon state in the Ising limit
and $|n\rangle$ represents
basis states (eigenstates in the Ising limit) which are not
degenerate with the single-magnon states.
However, the states $|\psi_i\rangle$ for different $i$ are
not orthogonal to each other when $\alpha\ne 0$.
Thus in order to construct the projection operator,
we need to define the overlap matrix
$g_{i,j}=<\psi_i|\psi_j>$.
Then the projection operator onto the single-magnon
subspace is given by the expression \cite{inverse}
\begin{equation}
{\cal P}=\sum_{i,j} g^{-1}_{i,j}|\psi_i><\psi_j|.
\end{equation}
The expansion coefficients for the residues in real space as
a function of the vector distance are given in Table~\ref{res_table}.
The coefficients for the transverse structure factor
are given in Ref.~5. 

We can now estimate the multi-magnon contribution to neutron
scattering by simply subtracting $A({\bf q})$ from the
equal-time correlation $\hat S({\bf q})$.
To get to the Heisenberg limit a series extrapolation is needed.
Since the series for
the multi-magnon weights are reduced by two terms (the first two
being zero) compared to the parent series, one might suspect
it would be better
to extrapolate the series for the total cross section and
the single-magnon contribution and take differences; we have
used both methods to estimate the multi-magnon weights,
carrying out the extrapolations by direct Pad\'e approximants.
In Fig.~\ref{weight_fig} results are presented for both the
multi-magnon spectral weight as well as the ratio of the multi-magnon
weight to the total spectral weight, along several lines in the
Brillouin zone.  We see that the multi-magnon contribution is
particularly large near $(\pi,0)$ (where, unfortunately, the
extrapolation uncertainties are largest as well) and amounts to
roughly a quarter of the total spectral weight.
The spin-wave calculations of Igarashi and Watabe\cite{IgarashiWatabe}
yield roughly twice as much spectral weight
in the multi-magnon excitations as we find by expansions
about the Heisenberg-Ising model; however, given Igarashi's later
remarks \cite{Igarashi} about the incorrect treatment of umklapp
processes in that work, we do not view the discrepancy as a
reason to question the accuracy of our results.

Recently, Stringari \cite{stringari}
has developed general bounds and sum rules
for single-magnon and multi-magnon spectra at special
wavevectors. Near ${\bf q}=(0,0)$ the
single magnon spectral weight vanishes linearly while the
multi-magnon weight vanishes quadratically; and
near  ${\bf q}=(\pi,\pi)$, the single-magnon spectral
weight diverges as $|(\pi,\pi)-{\bf q}|^{-1}$ while
the multi-magnon weight goes to a constant.
Our results are consistent with all of these requirements.

In summary, series expansions and extrapolations have
been carried out for dynamic properties of the $S=1/2$ square
lattice Heisenberg-Ising model.
Our numerical results indicate that the
Heisenberg model spin-wave spectrum is close to but noticeably
different from a uniformly renormalized classical spectrum.
In addition, the single-magnon and multi-magnon
spectral weights have been estimated throughout the Brillouin zone.
In light of our numerical results, it would be interesting to examine
the inelastic neutron scattering data on the antiferromagnetic parent
compounds of the cuprate superconductors to look for the multi-magnon
excitations.

{\it Acknowledgments.} We would like to thank Dr. G. Aeppli
for discussions. This work is supported by
National Science Foundation Grants DMR 93--18537 (RRPS)
and DMR 94--57928 (MPG).

\begin{figure}
\caption{Spin-wave spectrum for the Heisenberg model, in units of $J$,
along three lines in reciprocal space.  The solid circles with error
bars are the results from the series expansions; the solid line
is the classical spin-wave spectrum multiplied by an overall factor
$Z_c=1.18$.}
\label{disp_fig}
\end{figure}

\begin{figure}
\caption{Heisenberg model multi-magnon spectral weight (crosses,
narrowest error bars) and the ratio of the multi-magnon spectral
weight to the total spectral weight as determined by extrapolations
for the multi-magnon weight series (solid squares, widest error bars)
and by the difference of extrapolations for the total weight and
the single-magnon weight (diamonds, intermediate width error bars).}
\label{weight_fig}
\end{figure}

\begin{table}
\caption{Effective Hamiltonian for the Heisenberg-Ising model
elementary excitations in real space, up to an overall factor of four.
The dispersion in reciprocal
space is found by summing all of the given real-space series
with a factor $(1/4)[\cos(q_x r_x + q_y r_y) + \cos(q_x r_x - q_y r_y)
+ \cos(q_x r_y + q_y r_x) \break + \cos(q_x r_y - q_y r_x)]$, and
then dividing by four.}
\begin{tabular}{cl}
 ${\bf r}$ & Series \\
\hline
 (0,0) & $8 - 0.666662 \alpha^2  + 0.664352 \alpha^4$ \\
       & $-0.292737 \alpha^6 + 0.201076 \alpha^8 -0.177446
                                                   \alpha^{10}$ \\
 (1,1) & $-4 \alpha^2 + 1.222222 \alpha^4 - 0.541756 \alpha^6 $ \\
       & $+ 0.513359 \alpha^8 - 0.430609  \alpha^{10}$ \\
 (2,0) & $-2 \alpha^2 + 0.111111 \alpha^4 - 0.290209 \alpha^6 $ \\
       & $+ 0.255912 \alpha^8 -0.331992 \alpha^{10}$ \\
 (2,2) & $-0.291667 \alpha^4 -0.071979 \alpha^6 + 0.060564
                                                     \alpha^8 $ \\
       & $-0.133137 \alpha^{10}$ \\
 (3,1) & $-0.388889 \alpha^4 -0.177120 \alpha^6 + 0.143711
                                                     \alpha^8 $ \\
       & $-0.220432 \alpha^{10}$ \\
 (3,3) & $-0.072627 \alpha^6 + 0.002003 \alpha^8 -0.029703
                                                   \alpha^{10}$ \\
 (4,0) & $-0.048611 \alpha^4 - 0.074359 \alpha^6 + 0.021335
                                                     \alpha^8 $ \\
       & $-0.055154 \alpha^{10}$ \\
 (4,2) & $-0.108941 \alpha^6 - 0.013755 \alpha^8 -0.050071
                                                   \alpha^{10}$ \\
 (4,4) & $-0.016444 \alpha^8 - 0.010852 \alpha^{10}$ \\
 (5,1) & $-0.043576 \alpha^6 - 0.030875 \alpha^8 -0.034415
                                                   \alpha^{10}$ \\
 (5,3) & $-0.026311 \alpha^8 - 0.021359 \alpha^{10}$ \\
 (5,5) & $-0.005482 \alpha^{10}$ \\
 (6,0) & $-0.003631 \alpha^6 - 0.009438 \alpha^8 -0.011028
                                                   \alpha^{10}$ \\
 (6,2) & $-0.013155 \alpha^8 - 0.017717 \alpha^{10}$ \\
 (6,4) & $-0.009137 \alpha^{10}$ \\
 (7,1) & $-0.003759 \alpha^8 - 0.009967 \alpha^{10}$ \\
 (7,3) & $-0.005221 \alpha^{10}$ \\
 (8,0) & $-0.000235 \alpha^8 - 0.001596 \alpha^{10}$ \\
 (8,2) & $-0.001958 \alpha^{10}$ \\
 (9,1) & $-0.000435 \alpha^{10}$ \\
 (10,0) & $-0.000022 \alpha^{10}$ \\
\end{tabular}
\label{spec_table}
\end{table}

\begin{table}
\caption{Single-magnon spectral weight series in real space. To
evaluate the residue $A({\bf q})$ carry out the sum described
in the preceeding table caption.}
\begin{tabular}{cl}
 ${\bf r}$ & Series \\
\hline
 (0,0) & $0.5 - 0.041667 \alpha^2 + 0.011685 \alpha^4 $ \\
       & $-0.030642 \alpha^6 + 0.024677 \alpha^8 $ \\
 (1,0) & $-0.666667 \alpha + 0.110185 \alpha^3 -0.130751 \alpha^5 $ \\
       & $+0.117059 \alpha^7 -0.153300 \alpha^9 $ \\
 (1,1) & $0.194444 \alpha^2  + 0.084876 \alpha^4 - 0.035514
                                                         \alpha^6 $ \\
       & $+0.092202 \alpha^8 $ \\
 (2,0) & $0.097222 \alpha^2  + 0.072647 \alpha^4 - 0.015601
                                                         \alpha^6 $ \\
       & $+0.057647 \alpha^8 $ \\
 (2,1) & $-0.216667 \alpha^3 - 0.096930 \alpha^5 +0.055191
                                                         \alpha^7 $ \\
       & $-0.151490 \alpha^9 $ \\
 (2,2) & $0.065365 \alpha^4 + 0.008393 \alpha^6 + 0.030857
                                                         \alpha^8 $ \\
 (3,0) & $-0.036111 \alpha^3 - 0.049374 \alpha^5 +0.006892
                                                         \alpha^7 $ \\
       & $-0.045770 \alpha^9 $ \\
 (3,1) & $0.087153 \alpha^4 + 0.034750 \alpha^6 + 0.054776
                                                         \alpha^8 $ \\
 (3,2) & $-0.074005 \alpha^5 - 0.018980 \alpha^7 -0.049672
                                                         \alpha^9 $ \\
 (3,3) & $0.018793 \alpha^6 + 0.015430 \alpha^8 $ \\
 (4,0) & $0.010894 \alpha^4 + 0.018817 \alpha^6 + 0.018378
                                                         \alpha^8 $ \\
 (4,1) & $-0.037003 \alpha^5 - 0.028024 \alpha^7 -0.036609
                                                         \alpha^9 $ \\
 (4,2) & $0.028190 \alpha^6 + 0.029517 \alpha^8 $ \\
 (4,3) & $-0.023254 \alpha^7 -0.021256 \alpha^9 $ \\
 (4,4) & $0.007614 \alpha^8 $ \\
 (5,0) & $-0.003700 \alpha^5 - 0.009765 \alpha^7 -0.011880
                                                         \alpha^9 $ \\
 (5,1) & $0.011276 \alpha^6 + 0.021890 \alpha^8 $ \\
 (5,2) & $-0.013953 \alpha^7 -0.018825 \alpha^9 $ \\
 (5,3) & $0.012182 \alpha^8 $ \\
 (5,4) & $-0.009425 \alpha^9 $ \\
 (6,0) & $0.000940 \alpha^6 + 0.004772 \alpha^8 $ \\
 (6,1) & $-0.004651 \alpha^7 -0.011938 \alpha^9 $ \\
 (6,2) & $0.006091 \alpha^8 $ \\
 (6,3) & $-0.006284 \alpha^9 $ \\
 (7,0) & $-0.000332 \alpha^7 -0.002181 \alpha^9 $ \\
 (7,1) & $0.001740 \alpha^8 $ \\
 (7,2) & $-0.002693 \alpha^9 $ \\
 (8,0) & $0.000109 \alpha^8 $ \\
 (8,1) & $-0.000673 \alpha^9 $ \\
 (9,0) & $-0.000037 \alpha^9 $ \\
\end{tabular}
\label{res_table}
\end{table}


\begin{references}
\bibitem{inverse} The inverse of the matrix $g=I+X$,
where $I$ is the identity matrix and $X$ is of order $x$,
can be calculated perturbatively by
using the identity $g^{-1}=(I-X)(I+X^2)(I+X^4)(I+X^8)+ O(x^{16})$.
\end{references}
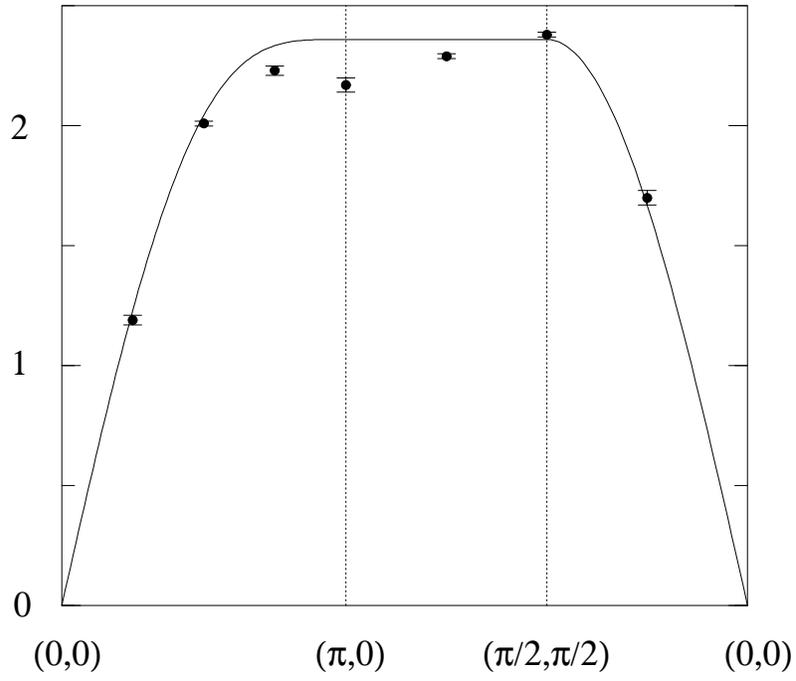
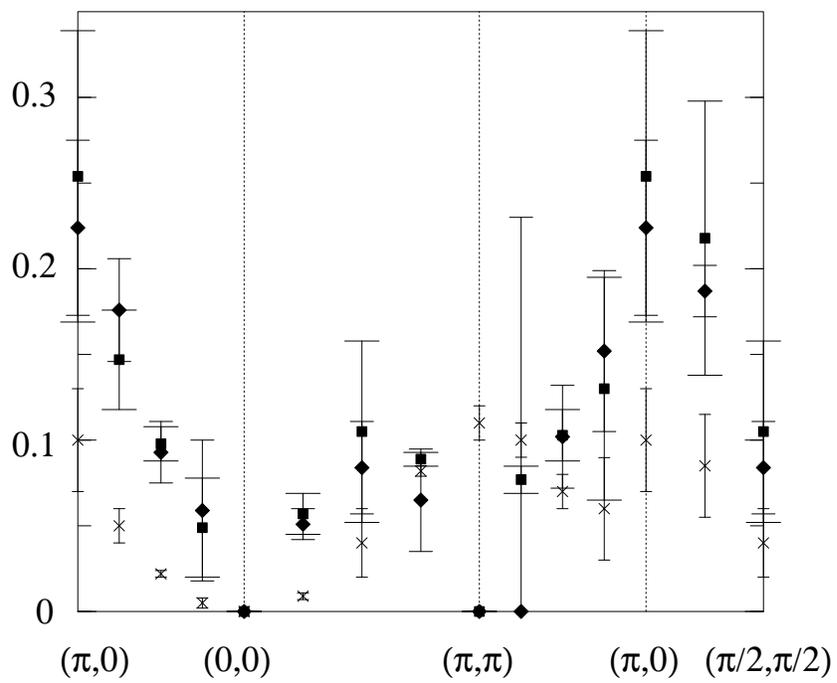
\end{document}